\documentclass[aps,prb,twocolumn,groupedaddress,superscriptaddress,nofootinbib]{revtex4-1}

\usepackage{graphicx}
\usepackage[colorlinks]{hyperref}
\hypersetup{
  colorlinks,
  citecolor=blue,
  linkcolor=red,
  urlcolor=blue}
\usepackage[english]{babel}
\usepackage{enumitem}
\usepackage{multirow}
\usepackage{array}
\usepackage{amsmath}
\newcolumntype{N}{>{\centering\arraybackslash}m{2cm}}
\newcolumntype{G}{>{\centering\arraybackslash}m{3cm}}

\begin{document}

%\preprint{}

\title{Micropillars with a controlled number of site-controlled quantum dots}


\author{Arsenty Kaganskiy}
\email[]{arsenty.kaganskiy@tu-berlin.de}
%\thanks{}
\affiliation{Institut f{\"u}r Festk{\"o}rperphysik, Technische Universit{\"a}t Berlin, Hardenbergstra{\ss}e 36, D-10623 Berlin, Germany}

\author{Fabian Gericke}
%\thanks{}
\affiliation{Institut f{\"u}r Festk{\"o}rperphysik, Technische Universit{\"a}t Berlin, Hardenbergstra{\ss}e 36, D-10623 Berlin, Germany}

\author{Tobias Heuser}
%\thanks{}
\affiliation{Institut f{\"u}r Festk{\"o}rperphysik, Technische Universit{\"a}t Berlin, Hardenbergstra{\ss}e 36, D-10623 Berlin, Germany}

\author{Tobias Heindel}
%\thanks{}
\affiliation{Institut f{\"u}r Festk{\"o}rperphysik, Technische Universit{\"a}t Berlin, Hardenbergstra{\ss}e 36, D-10623 Berlin, Germany}

\author{Xavier Porte}
%\thanks{}
\address{Institut f{\"u}r Festk{\"o}rperphysik, Technische Universit{\"a}t Berlin, Hardenbergstra{\ss}e 36, D-10623 Berlin, Germany}

%\author{Andr\'{e} Strittmatter}
%\thanks{}
%\affiliation{Institut f{\"u}r Festk{\"o}rperphysik, Technische Universit{\"a}t Berlin, %Hardenbergstra{\ss}e 36, D-10623 Berlin, Germany}
%\affiliation{Abteilung Halbleiterepitaxie, Otto-von-Guericke Universit{\"a}t Magdeburg,  %Universit{\"a}tsplatz 2, D-39106 Magdeburg, Germany}

\author{Stephan Reitzenstein}
%\thanks{}
\affiliation{Institut f{\"u}r Festk{\"o}rperphysik, Technische Universit{\"a}t Berlin, Hardenbergstra{\ss}e 36, D-10623 Berlin, Germany}


\date{\today}

\clearpage

\begin{abstract}
We report on the realization of micropillars with site-controlled quantum dots (SCQDs) in the active layer. The SCQDs are grown via the buried stressor approach which allows for the positioned growth and device integration of a controllable number of QDs with high optical quality. This concept is very powerful as the number and the position of SCQDs in the cavity can be simultaneously controlled by the design of the buried stressor. The fabricated micropillars exhibit a high degree of position control for the QDs above the buried stressor and $Q$-factors of up to 12000 at an emission wavelength around 930~nm. We experimentally analyze and numerically model the cavity $Q$-factor, the mode volume, the Purcell factor and the photon-extraction efficiency as a function of the aperture diameter of the buried stressor. Exploiting these SCQD micropillars, we experimentally observe the Purcell enhancement in the single-QD regime with $F\textsubscript{P}$~=~4.3~$\pm$~0.3. 
\end{abstract}

%\pacs{}
%\keywords{}

\maketitle

\section{Introduction}

%acceleratesfor fabrication of various semiconductor devices based on integrated light emitters into optical microresonators.\citep{Aharonovich.2016}
%  They have already been proved as efficient light emitters in photonic wires,\citep{Reimer.2012} photonic crystals\citep{Madsen.2014} and micropillar cavities.\citep{Unsleber.2016}

%The increase of the $\beta$-factor leads to beneficial effects resulting in a boosting of the outcoupling efficiency of SPSs as well as in a reduction of the threshold of microlasers.

The enormous progress in the development of nanofabrication technologies and nanophotonic device concepts has boosted experimental and theoretical studies in the field of cavity quantum electrodynamics (cQED).\citep{Gerard.1998, Noda.2007} A variety of devices exploiting cQED effects has been developed in recent years, including highly-efficient single-photon sources (SPSs)\citep{Michler.2000, Heindel.2010, Gazzano.2013, Schlehahn.2016, Ding.2016, Unsleber.2016} for the realization of, e.g., quantum key distribution\citep{Waks.2002, Heindel.2012}, high-$\beta$\citep{Strauf.2006, Prieto.2015, Reitzenstein.2008, Reitzenstein.2008b} as well as thresholdless lasers.\citep{Ota.2017, Khajavikhan.2012} Furthermore, the coexistence of lasing and strong coupling has already been demonstrated in nanocavities.\citep{Nomura.2010, Gies.2017} An attractive type of system for the exploitation of cQED effects are micropillar cavities with integrated semiconductor quantum dots (QDs).\citep{Reitzenstein.2010} Here, the huge interest in semiconductor QDs is explained by their excellent optical properties combined with the scalability of the solid-state host system. Noteworthy, the spectral and spatial matching of the emitters to the cavity mode needs to be ensured in order to maximize the achievable Purcell factor $F\textsubscript{P}$\citep{Gerard.1998} and the related spontaneous emission factor $\beta=F\textsubscript{P}$/$(F\textsubscript{P}+1)$. The latter describes the fraction of spontaneous emission funneled into the cavity mode and is crucial for the optimization of microcavity based quantum light sources and lasers.  Common self-organized growth-modes for QDs result in a random spatial and spectral (within the inhomogeneous broadening of the QD ensemble) position which hinders a controlled fabrication of the aforementioned devices. Different strategies have been developed to overcome this issue via the deterministic integration of QDs into resonators. These strategies can be divided in two groups: a) concepts based on postgrowth alignment of resonators to selected self-assembled grown QDs via $in-situ$ lithography approaches\citep{Dousse.2008} and b) concepts based on the site-controlled growth of QDs.\citep{Schneider.2008} Though the first approach has been very successfully applied in recent years it is limited to the realization of single-QD devices such as SPSs. The site-controlled approach is technologically more demanding. On the one hand, it allows for true scalability by the fabrication of ordered arrays of emitters.\citep{Surrente.2009} On the other hand, the approaches used to enforce the site-controlled growth typically lead to a reduced optical quality of the emitter reflecting a close proximity of QDs to the etched surface.\citep{Albert.2010} In order to overcome this challenge structures with several stacked QD layers have been realized\citep{Schneider.2009} demonstrating SCQDs with a high optical quality. However, this approach is still limited to a minimum pitch of 200~nm between single QDs.\citep{Huggenberger.2011} In contrast, high-$\beta$ microlasers would strongly benefit from the integration of QD-ensembles consisting of a defined number of emitters located at the anti-node of the laser mode with a lateral extension of a few hundred nm. In this regard, the buried-stressor growth approach based on a partial oxidation of AlAs aperture is highly interesting. It leads to the positioned growth of QDs aligned to a buried oxide-aperture where the number of QDs can be controlled by the aperture diameter.\citep{Strittmatter.2012} In the limit of small diameters it has already been applied for the fabrication of high-quality optically and electrically driven SPSs.\citep{Strau.2017, Unrau.2012, Kaganskiy.2017}

In this work, we apply the buried-stressor growth concept with oxidized apertures to realize high-quality micropillar structures with a controllable number of SCQDs as gain medium. We fabricate micropillars with about 1 to 20 SCQDs positioned close the maximum of the fundamental cavity mode. This specific configuration of the gain-medium is not feasible with any other growth and processing scheme and is highly attractive for the fabrication of SPSs and microlasers operating in the few-QD regime. The present work focuses on the development of the required growth and processing technology and a detailed optical investigation of the realized SCQD micropillar cavities. It includes a systematic study of the micropillars in terms of the number of SCQDs, emission energy, effective mode volume and quality $(Q)$ factor as a function of the diameter of the oxide-aperture. Using our novel type of micropillars, we are able to observe the Purcell enhancement of a single QD.
\\
\section{Methods}
\label{sec:methods}

\begin{figure}[t!]
\centering
\includegraphics[width=8 cm]{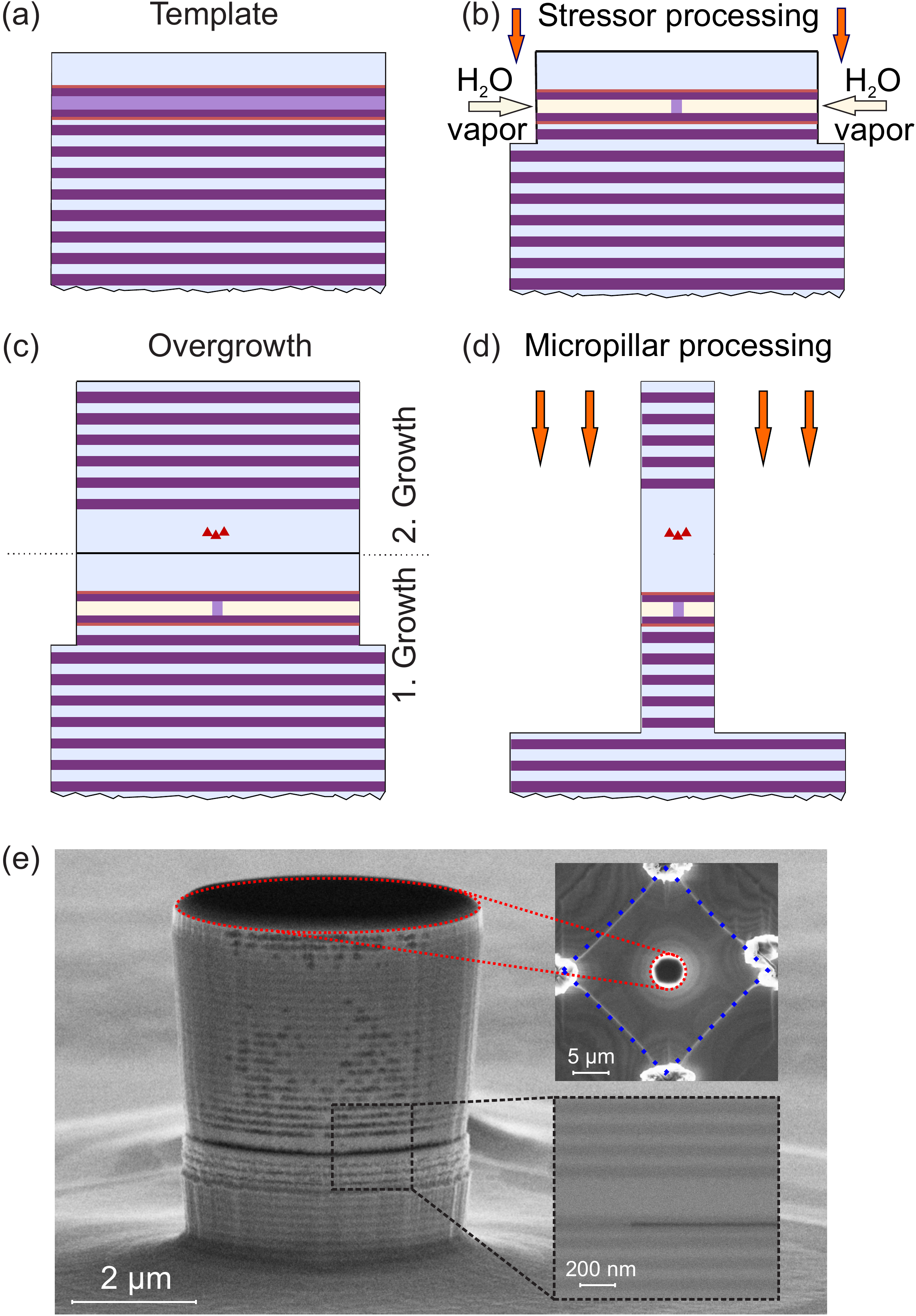}
\caption{(a-d) Schematic illustration of SCQD growth and patterning of micropillars via the buried stressor approach. (a) The growth of a template structure is followed by the mesa processing and the subsequent oxidation of the AlAs layer acting as buried stressor (b). (c) SCQDs located in the center of a $\lambda$-cavity and the top DBR are grown in the second MOCVD step. (d) The fabrication is finalized by  micropillar processing via EBL and ICP-RIE dry etching. (e) SEM image of a fully processed micropillar structure. Insets: (top) top-view of the square mesa (marked with a blue dashed rhombus) processed in the step (b) with a micropillar (marked with a red dashed circle) aligned to its center. (bottom) Cross-sectional SEM image of the bottom and top DBR and of the partially oxidized AlAs layer.}
\label{fig:fabrication}
\end{figure}

The work flow for the fabrication of micropillars comprises two-step metal-organic chemical vapor deposition (MOCVD) epitaxial growth and cleanroom processing as schematically shown in Fig.~\ref{fig:fabrication}~(a~-~d). Firstly, a so-called template structure is grown at 700~$^\circ$C on an n-doped GaAs substrate. It includes the lower distributed Bragg reflector (DBR) and a layer sequence consisting of a 30~nm-thick AlAs layer embedded into 40~nm-thick Al$\textsubscript{0.90}$Ga$\textsubscript{0.10}$As claddings and 10~nm thick strain reducing AlGaAs layers (Fig.~\ref{fig:fabrication}~(a)). The first growth is completed by an 80~nm-thick GaAs layer representing a bottom part of the $\lambda$-cavity. The DBR mirror consists of 27~pairs of $\lambda$/4-thick Al$\textsubscript{0.90}$Ga$\textsubscript{0.10}$As/GaAs-layers in which the thickness of the last GaAs layer is reduced to 22~nm in order to align the AlAs layer to an antinode of the electrical field. Next, arrays of square mesas with a side length varying between 20.4~and~21.4~$\mu$m (step-size 67~nm) are patterned via optical lithography and dry etched down to the uppermost mirror-pair of the lower DBR in order to access the AlAs layer for the further lateral oxidation (cf. Fig.~\ref{fig:fabrication}~(b)) which is subsequently performed under H$\textsubscript{2}$O/N$\textsubscript{2}$ atmosphere at 420~$^{\circ}$C. Control of the aperture diameter with an accuracy of a few hundred nanometers is achieved by $in-situ$ optical monitoring during the oxidation process. Prior to the second epitaxial growth step the sample is dipped into a 75~$\%$ sulfuric acid in order to remove oxides from the surface. This cleaning step is crucial for the realization of a defect-free overgrowth of the $\approx$~3.5~$\mu$m thick upper DBR. The MOCVD overgrowth starts with a 50~nm-thick GaAs buffer layer grown at 700~$^\circ$C followed by In$\textsubscript{0.60}$Ga$\textsubscript{0.40}$As QDs grown in Stranski-Krastanov growth mode at 500~$^\circ$C. Afterwards, the upper half of the GaAs cavity with a thickness of 130~nm is grown at 615~$^\circ$C. The growth is finalized by the upper DBR mirror consisting of 23~pairs of $\lambda$/4 thick Al$\textsubscript{0.90}$Ga$\textsubscript{0.10}$As/GaAs-layers grown at 615~$^\circ$C (cf. Fig.~\ref{fig:fabrication}~(c)). After the overgrowth, a 300~nm-thick Si$\textsubscript{3}$N$\textsubscript{4}$ layer is deposited on the sample by plasma-enhanced chemical vapor deposition (PECVD) followed by spin-coating of a 400~nm thick layer of a negative resist AZ~nLOF~2070. Then micropillar structures are defined via electron beam lithography (EBL) and realized by inductively-coupled-plasma reactive-ion dry etching (ICP-RIE) (Fig.~\ref{fig:fabrication}~(d)). Here, the micropillars are aligned to the center of the square mesas (corresponding to the oxide-apertures) and the associated SCQDs with high position accuracy of about 200~nm (cf. top inset of Fig.~\ref{fig:fabrication}~(e)). Fig.~\ref{fig:fabrication}~(e) shows a scanning electron microscope (SEM) image of a fully processed micropillar with a diameter of 4.4~$\mu$m. The microcavity structure is etched down to the 17th pair of the bottom DBR and the realized micropillar has almost vertical sidewalls with low surface roughness. The inset of Fig.~\ref{fig:fabrication}~(e) presents a cross-sectional zoom-in view of the central part of the structure before the last dry etching step including the partially oxidized AlAs (transition from the grey AlAs to the black oxidized layer). It also shows high-quality defect-free epitaxial layers of the cavity structure.    

\section{Results and Discussion}

\begin{figure}[t!]
\centering
\includegraphics[width=8 cm]{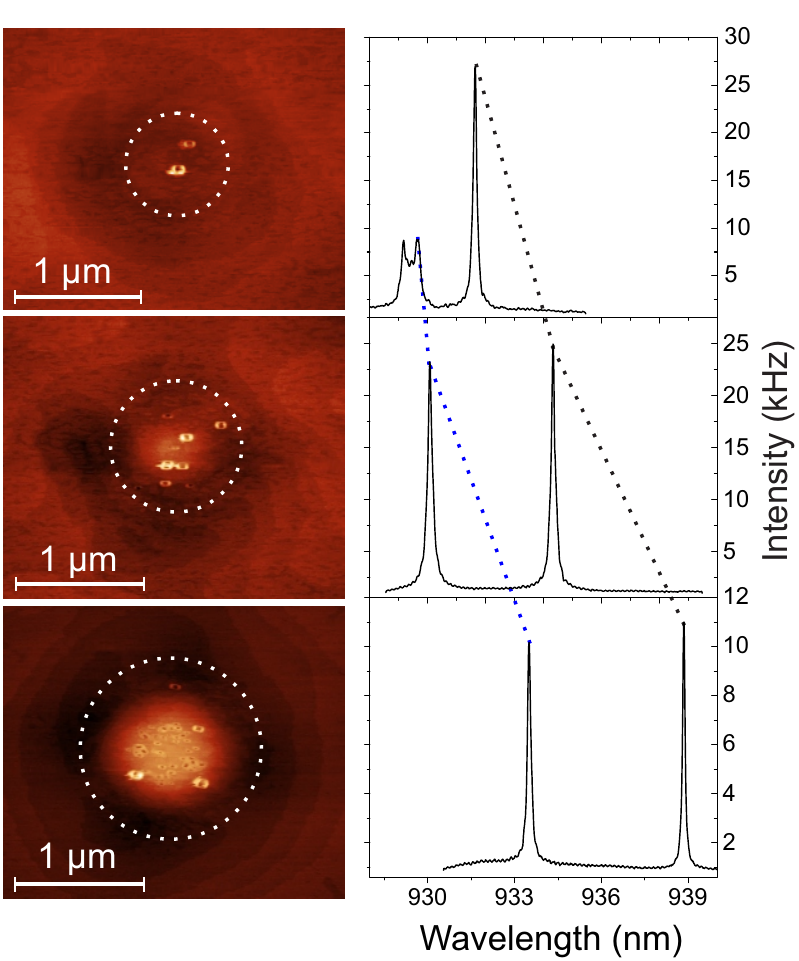}
\caption{Left column: Atomic-force microscope (AFM) images of a reference sample surface after the overgrowth with SCQDs demonstrating that the number of positioned QDs increases with the aperture diameter which is marked with a white dashed circle. Right column: $\mu$PL spectra of micropillars with the identical aperture size (similar number of SCQDs) than their left column counterparts and a pillar diameter of 4.4~$\mu$m. Spectral shifts of the fundamental and the first higher order transverse mode for different aperture diameters are indicated by dashed lines and are associated with an aperture dependent mode confinement.}
\label{fig:AFM-Spectra}
\end{figure}

\begin{figure}[t!]
\centering
\includegraphics[width=8 cm]{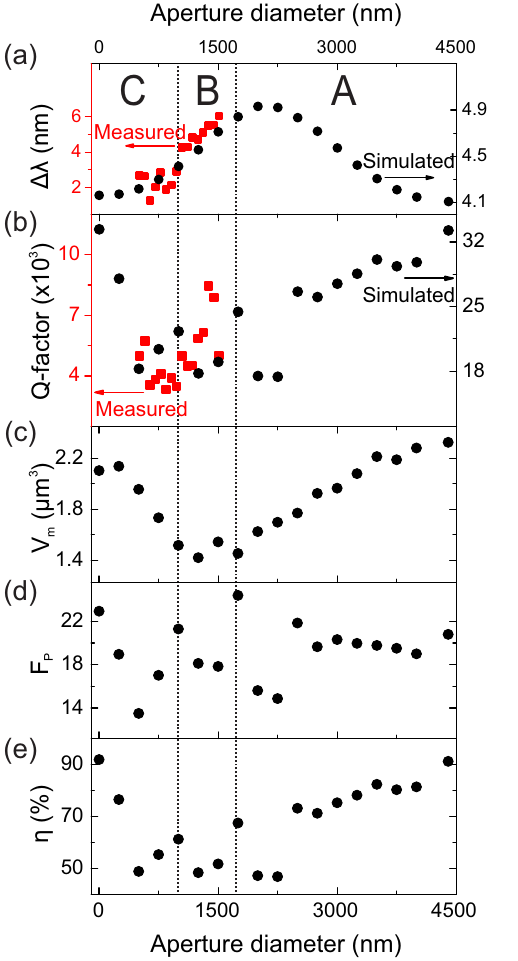}
\caption{Numerically simulated (black circles, labeled on the right column) and experimentally measured (red squares, left column) mode splitting $\Delta\lambda$ (a) and $Q$-factor (b) as a function of the aperture diameter. Simulated mode volume $V\textsubscript{m}$ (c), Purcell factor $F\textsubscript{P}$ (d) and photon-extraction efficiency $\eta$ (e) as a function of the aperture diameter. All results correspond to the micropillars with the design presented in Sec.~\ref{sec:methods} with a diameter of 4.4~$\mu$m. All plots include sub-panels as described in the text.}
\label{fig:Simulations-Modes}
\end{figure}

The fabricated micropillar structures are investigated via micro-photoluminescence ($\mu$PL) spectroscopy at 10 - 50 K. The sample is mounted inside a He-flow cryostat and the investigated micropillars are optically excited by a continuous-wave diode laser emitting at a wavelength of 671~nm. The resulting photoluminescence is collected via a microscope objective with a numerical aperture of 0.4, spectrally dispersed by a spectrometer including a Si-based charge-coupled device camera with an overall 
spectral resolution of 25~$\mu$eV.

In order to demonstrate and investigate the site-control of our QD-growth technique, we prepared a reference sample nominally identical to the sample described above, but without upper DBR and with a 3~nm-thick GaAs capping layer above the SCQDs for atomic-force microscopy characterization. The results are presented in the left column of Fig.~\ref{fig:AFM-Spectra}. As expected,\citep{Strittmatter.2012} the number of QDs increases with increasing aperture diameter starting from 2 QDs for an aperture diameter of 700~nm (top image), leading to 9 QDs for an aperture of 1000~nm and resulting in a small QD-ensemble for an aperture of 1400~nm. Noteworthy, there is no QD-growth apart from the aforementioned SCQDs showing a high selectivity of our growth technique. On the right column of Fig.~\ref{fig:AFM-Spectra} $\mu$PL spectra of processed micropillars with a diameter of 4.4~$\mu$m and with the nominally same aperture diameters as in the left column are presented. We observe emission of the fundamental pillar mode on the low-energy side and of the first higher transverse mode on the high-energy side, respectively. Interestingly, for the given pillar diameter both, the emission energy and the mode splitting depend on the aperture diameter which we attribute to its influence on the mode volume of the micropillar cavity. The oxide-aperture influence on the emission features is investigated experimentally for a family of micropillars with constant diameter of 4.4~$\mu$m and aperture diameters in the range of 500 to 1500~nm. Additionally, numerical simulations are performed for aperture diameters in the range from 0 to 4400~nm by using the finite-element software package JCMsuite\citep{JCMwaveGmbH.} for solving second order Maxwell's equations of the electric field.\citep{Karl.2009} The results of these studies are presented in Fig.~\ref{fig:Simulations-Modes} where different parameters, such as the mode splitting $\Delta\lambda$, the quality factor $Q$, the mode volume $V\textsubscript{m}$, the Purcell factor $F\textsubscript{P}$ and the photon-extraction efficiency $\eta$ are depicted as a function of the aperture diameter. We subdivide the presented results into three characteristic diameter ranges separated by dashed vertical lines (cf. Fig.~\ref{fig:Simulations-Modes}). The simulation results of the mode behavior for the aperture diameters between 1700 and 4400~nm (region A) are in agreement with previous works\citep{Bakker.2013, Bennett.2007} and show an increase of the mode splitting $\Delta\lambda$ (Fig.~\ref{fig:Simulations-Modes}~(a)) and a reduction of the mode volume $V\textsubscript{m}$ (Fig.~\ref{fig:Simulations-Modes}~(c)) with a decreasing aperture diameter, which is explained by enhanced mode confinement. The latter leads to a blue-shift of the energy of the fundamental mode\citep{Ellis.2008} (not shown here). Due to light scattering induced by the refractive index contrast between the oxidized and non-oxidized material, the $Q$-factor decreases with  decreasing aperture diameter (Fig.~\ref{fig:Simulations-Modes}~(b)).\citep{Weidenfeld.2013} In contrast, the Purcell factor $F\textsubscript{P}$ does not show a systematic dependence on the oxide-aperture diameter (Fig.~\ref{fig:Simulations-Modes}~(d)) by mutually compensating the influence of the decreasing $V\textsubscript{m}$ and $Q$. 

\begin{figure}[t!]
\centering
\includegraphics[width=8 cm]{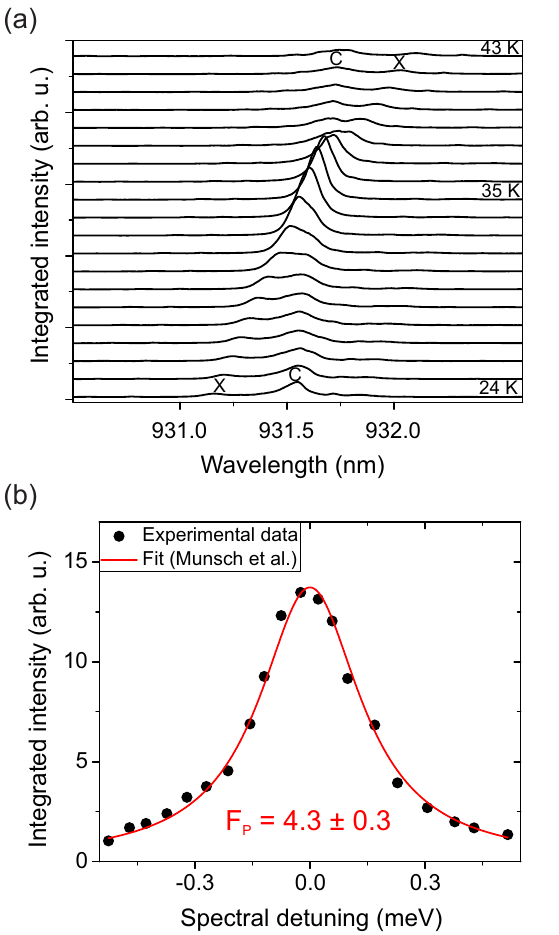}
\caption{(a) Waterfall presentation of temperature dependent $\mu$PL spectra showing the spectral tuning of a single-QD exciton (X) through resonance with the fundamental cavity mode (C) of a 4.4~$\mu$m micropillar with oxid-aperture diameter of 1300~nm and a Q-factor of 8000. cQED enhanced light-matter interaction results in pronounced Purcell enhancement at the resonance temperature of 35~K. (b) Integrated intensity of the QD emission as a function of the spectral detuning between X and C. The fit according to Munsch et al.\citep{Munsch.2009} (red line) yields $F\textsubscript{P}$~=~4.3~$\pm$~0.3.}
\label{fig:Purcell}
\end{figure}

For the diameter regions B and C the theoretical findings are compared with the experimental data. In the intermediate region B corresponding to aperture diameters between 800 and 1700~nm, the mode confinement starts to decrease and $\Delta\lambda$ decreases accordingly. Here, the scattering-induced reduction of the experimental $Q$-factor is confirmed qualitatively by the simulation and results in a reduced photon-extraction efficiency $\eta$ (cf. Fig.~\ref{fig:Simulations-Modes}~(e)), which is expressed as $\frac{Q\textsubscript{MP}}{Q\textsubscript{0}}\frac{F\textsubscript{P}}{F\textsubscript{P}+1}$,\citep{Barnes.2002} where $Q\textsubscript{MP}$ and $Q\textsubscript{0}$ are the Q-factor of the micropillar and the planar Q-factor, respectively. In region C the shrinking and possibly closing aperture leads to an increase of $V\textsubscript{m}$ as well as of $Q$ and, hence, of $\eta$. This observation is explained by a decoupling of the mode from the refractive index contrast associated with oxidized AlAs when the aperture diameter becomes smaller than the lateral mode extension. The deviation of the measured $Q$-factors from the values predicted by simulations is attributed partly to absorption losses, scattering losses at layer interfaces of the overgrown microcavity and photon losses at the micropillar's sidewalls, which are not considered in the simulations. 
% Calculated of micropillars with a constant aperture diameter of 700~nm (not presented here) shows a continuous increase of the $Q$-factor with the increasing micropillar diameter which is experimentally approved resulting in a $Q$-factor of 6000 and 12000 for micropillars with a diameter of 4 and 6.5~$\mu$m, respectively.

In order to demonstrate pronounced single-QD cQED effects and to confirm the predicted enhancement of the spontaneous emission rate (cf.~Fig.~\ref{fig:Simulations-Modes}~(d)), we performed temperature-tuning of a single-QD exciton (X) through the fundamental cavity mode (C) of a micropillar with a diameter of 4.4~$\mu$m and an aperture diameter of 1.3~$\mu$m. Fig.~\ref{fig:Purcell}~(a) presents the corresponding $\mu$PL spectra in the temperature range between between 24 and 43~K. At resonance at 35~K we observe pronounced enhancement of QD-X emission due to the Purcell effect. In order to determine the associated Purcell factor $F\textsubscript{P}$ we extracted the detuning dependent integrated intensity of X emission by Lorentzian lineshape fitting. The corresponding data is plotted in Fig.~\ref{fig:Purcell}~(b) as a function of the spectral detuning $\Delta$ between X and C. By fitting this dependence via $I(\Delta)~=~F\textsubscript{P}/(1~+~F\textsubscript{P}~+~4\Delta^2/\gamma\textsubscript{C}^2)$\citep{Munsch.2009}, where $\gamma\textsubscript{C}$ is the cavity linewidth, we determine $F\textsubscript{P}$~=~4.3~$\pm$~0.3. This value is lower in comparison to the simulated value of $F\textsubscript{P}$ = 18 for an ideal structure. Taking the experimental Q-factor into account, the upper limit of the Purcell factor can be calculated via $F\textsubscript{P}~=~(3Q\textsubscript{exp}/4\pi^2V\textsubscript{M})(\lambda^3/n\textsubscript{eff}^3)$\citep{Gerard.1998} which yields $F\textsubscript{P}\textsuperscript{max}$~=~11. This value is still higher than the experimental $F\textsubscript{P}$ which we attribute to a possible polarization mismatch between X and C and to a non-ideal spatial matching between the QD and the maximum of the fundamental micropillar mode.\citep{Kaganskiy.2017} 

\section{Summary and Conclusion}

In conclusion we presented an attractive fabrication platform for the realization of micropillar cavities with a controlled number of SCQDs in the active layer. The structures are fabricated by the buried-stressor growth concept and studied comprehensively in terms of the influence of the buried oxide-aperture on the optical properties of the micropillars. Both the $Q$-factor and the mode volume depend in a characteristic way on the diameter of the oxide-aperture. It leads to additional mode-confinement and scattering for a given micropillar diameter, thereby influencing the mode volume, the cavity Q-factor and the associated Purcell-factor and photon-extraction efficiency. The fabricated micropillars exhibit $Q$-factors of up to 12000 and a pronounced single-QD Purcell enhancement with $F\textsubscript{P}$~=~4.3~$\pm$~0.3. The demonstrated approach is well suited for the realization of cQED-enhanced SPSs and paves the way for the development of high-$\beta$ microlasers with a controlled number of SCQDs aligned spatially to the laser mode.     
 
\section{Acknowledgments}

The research leading to these results received funding from the European Research Council under the European Union\textsc{\char13}s Seventh Framework Program Grant Agreement No. 615613, from the Volkswagen Foundation via NeuroQNet and from the German Research Foundation via CRC 787.

\end{document}